\documentstyle[12pt]{article}

    \let\b=\beta     \let\g=\gamma    \let\d=\delta
     \let\e=\epsilon
            \let\q=\theta       
 \let\k=\kappa
\let\l=\lambda                    \let\x=\xi
 \let\p=\pi
\let\r=\rho                      
   \let\f=\phi
\let\c=\chi               \let\vq=\vartheta       
\let\w=\omega

\let\X=\Xi                        
 
\let\Y=\Psi              \let\D=\Delta     \let\L=\Lambda

\let\la=\label

      \let\pa=\partial      \let\bm=\bibitem
\newcommand{\be}{\begin{equation}}
\newcommand{\ee}[1]{\label{#1}\end{equation}}
\newcommand{\bea}{\begin{eqnarray}}
\newcommand{\eea}{\end{eqnarray}}
\newcommand{\ra}{\rightarrow}

\newcommand{\dV}[1]{\frac{\d}{\d V(#1)}}
\newcommand{\dY}[1]{\frac{\d}{\d \Y (#1)}}
\newcommand{\sect}[1]{\setcounter{equation}{0} \section{#1}}
\renewcommand{\theequation}{\thesection .\arabic{equation}}
\newcommand{\refer}[1]{(\ref{#1})}
\newcommand{\Blangle}{\Bigl \langle}
\newcommand{\Brangle}{\Bigr \rangle}
\newcommand{\cint}[2]{\oint_{#2} \, \frac{d #1}{2\p i}}
\newcommand{\XL}{( \X_1 - \L_1 )}
\newcommand{\Yop}{\Bigl (\, \widehat{\Y} \, -\, v_0(p)\, \Bigr )}
\newcommand{\Vop}{\Bigl (\, \widehat{V}^\prime \, -\, u_0(p)\, \Bigr )}
\newcommand{\px}[1]{\frac{1}{(p-x)^{#1}}}
\newcommand{\py}[1]{\frac{1}{(p-y)^{#1}}}
\newcommand{\lra}{\leftrightarrow}
\newcommand{\bra}[1]{\Bigl [ \, #1 \, \Bigr ]}
\newcommand{\mat}[1]{\mbox{$ #1 $}}
\newcommand{\Fbg}{\mat{F^{\mbox{\scriptsize bos}}_g}\  }
\newcommand{\Ffg}{\mat{F^{\mbox{\scriptsize ferm}}_g}\  }

\begin{document}
\thispagestyle{empty}
{\hbox to\hsize{
\vbox{\noindent ITP--UH--04--95 \hfill January 1995 \\
hep-th/9501120}}}
\noindent
\vskip2cm
\begin{center}

{\Large\bf Iterative Solution of the Supereigenvalue Model}
\vglue2cm

Jan C.\ Plef\/ka \footnote{Supported by
the `Studienstiftung des Deutschen Volkes'}
\vglue1cm
{\it Institut f\"{u}r Theoretische Physik, Universit\"{a}t Hannover}\\
{\it Appelstra\ss{}e 2, 30167 Hannover, Germany}\\
{\footnotesize plefka@itp.uni-hannover.de}

\vglue2cm
{\large  ABSTRACT}
\end{center}
\noindent
An integral form of the discrete superloop equations for the supereigenvalue
model of Alvarez--Gaum\'e, Itoyama, Ma\~nes and Zadra is given. By a change
of variables from coupling constants to moments we find a compact form
of the planar solution for general potentials. In this framework an iterative
scheme for the calculation of higher genera contributions to the free energy
and the multi--loop correlators is developed. We present explicit results for
genus one.
\vfill
\setcounter{page}{0}
\pagebreak

\sect{Introduction}

After the successful application of matrix models to 2d gravity and
bosonic string theory \cite{2dGra}, it was natural to ask what may
be done for the supersymmetric case. Unfortunately the
description of discretized two--dimensional random surfaces by the hermitian
matrix model to date has no analogue in terms of a supersymmetric matrix
model describing a discretization of super--Riemann surfaces.

Nevertheless the supereigenvalue model proposed by Alvarez-Gaum\' e, Itoyama,
Ma\~ nes and Zadra \cite{Alv} appears to have all the virtues of a discrete
approach to 2d supergravity. Guided by the prominent role the Virasoro
constraints \cite{Vir} played for the hermitian matrix model, the authors
constructed a partition function built out of $N$ Grassmann even and odd
variables (the ``supereigenvalues'') as well as even and odd coupling
constants obeying a set of super--Virasoro constraints. As it is unknown
whether there exists a matrix--based formulation of this model, we do not have
a geometric interpretation of it at hand. However, the study of the
 supereigenvalue
model revealed that in its continuum limit it describes 2d supergravity
coupled
to minimal superconformal field theories \cite{Alv,Alv2}. Moreover, a precise
dictionary between continuum \mbox{$N=1$} super--Liouville amplitudes and
supereigenvalue correlators has been developed \cite{Zad}.

Just as the hermitian matrix model, the supereigenvalue model admits an
expansion in \mat{1/N^2}, termed the genus expansion.
The super--Virasoro constraints of the supereigenvalue model are equivalent
to a set of superloop equations for the superloop correlators, generalizing the
well--known loop equations of the hermitian matrix model \cite{Mig,loop,Mak}.
The superloop equations were solved in the planar \mat{N\ra\infty} limit
for general potentials away from the double scaling limit in \cite{Alv2}.
Higher genera results were obtained
in the double scaling limit. An alternative
approach was pursued by the authors of \cite{Bec,McA}, who managed to
directly integrate out the Grassmann--odd variables on the level of the
partition function. Interestingly enough this analysis revealed that the free
energy of the supereigenvalue model depends at most quadratically on the
fermionic coupling constants. Moreover, Becker and Becker \cite{Bec} showed
that this free energy is related in a simple way to the free energy of the
hermitian matrix model.

In this paper we shall be interested in the supereigenvalue model away from
the double scaling limit. We develop an iterative procedure to solve the
superloop equations for general potentials genus by genus. It represents a
generalization of the very effective iterative scheme of Ambj\o rn, Chekhov,
Kristjansen and Makeenko \cite{Amb} to calculate higher genus contributions
to the multi--loop correlators and the free energy of the hermitian one matrix
model. Our approach is based on a new integral form of the superloop equations
and the introduction of superloop insertion operators. Similarly to \cite{Amb}
the key point in this scheme is the change of variables from the coupling
constants to the so-called moments of the bosonic and fermionic potentials.
The advantage of these new variables is that a dependence on an infinite
number of coupling constants arranges itself nicely into a finite number of
moments at each genus.
With these methods at hand, we derive a new quite compact form of the
solution of the superloop equations in the planar limit. In addition
we present explicit
results for the free energy and the one--superloop correlators at genus one.
Moreover, it is shown that the genus $g$ contribution to the free energy
depends
on \mat{2(3\, g)} bosonic and \mat{2(3\, g+1)} fermionic moments.

The paper is organized as follows: In section 2 we introduce the basic
ingredients of our iterative solution, the superloop
insertion operators and the superloop equations. Section 3 then contains the
planar solution. A set of basis functions and our new
variables -- the moments -- are introduced. The iterative procedure
is developed
in Section 4. In parallel we present explicit results for genus one. Moreover
we state some details on the general structure of the free energy and the
superloop correlators at genus $g$. Finally in section 6 we conclude.

\sect{Superloop Equations}

Our solution of the supereigenvalue model proposed by Alvarez--Gaum\'e,
Itoyama, Ma\~nes and Zadra  \cite{Alv} is based on an integral form of the
superloop equations. Generalizing the approach of Ambj\o rn, Chekhov,
Kristjansen and Makeenko \cite{Amb}
for the hermitian one--matrix model we develop an iterative procedure which
allows us to calculate the genus $g$ contribution to the $(n|m)$-superloop
correlators for (in principle) any $g$ and any $(n|m)$ and (in practice) for
any potential. The possibility of going to arbitrarily high genus is
provided by
the superloop equations, whereas the possibility of obtaining arbitrary
$(n|m)$-superloop results is due to the superloop insertion operators
introduced below. A change of variables from the coupling constants to moments
allows us to explicitly present results for arbitrary potentials.

\subsection{Superloop Insertion Operators}

The supereigenvalue model \cite{Alv} is built out of
a set of $N$ bosonic and fermionic variables, denoted by $\l_i$ and $\q_i$
respectively. $N$ is even. The partition function is given by

\be
{\cal Z}= e^{N^2 \, F} =
\int (\prod_{i=1}^{N} d\l_i \, d\q_i ) \, \prod_{i<j}\, (\l_i -
\l_j - \q_i\q_j )
\,\exp \Bigl ( - N
\sum_{i=1}^N \, [ V(\l_i) - \q_i \Y (\l_i)] \, \Bigr )
\ee{model}
where

\be
V(\l_i)= \sum_{k=0}^\infty g_k {\l_i}^k   \quad \mbox{and} \quad \,\,\,
\Y(\l_i) = \sum_{k=0}^\infty \x_{k+1/2}  {\l_i}^k,
\ee{}
the $g_k$ and $\x_{k+1/2}$
being Grassmann even and odd coupling constants, respectively.
This model obeys a set of
super--Virasoro constraints $L_k \, {\cal Z}=0$ and $G_{k-1/2}\,
{\cal Z}=0$ for
$k \geq 0$ by construction. Here the super--Virasoro operators
are represented as differential operators in the coupling constants
$g_k$ and $\x_{k+1/2}$ \cite{Alv}.

Expectation values are defined in the usual way by

\be
\langle {\cal O}(\l_j,\q_j) \rangle = \frac{1}{\cal Z} \int (\prod_{i=1}^{N}
d\l_i \, d\q_i )\, \,\D (\l_i, \q_i ) \,\, {\cal O} (\l_j ,\q_j) \, \,\exp
\Bigl ( - N
\sum_{i=1}^N [ \, V(\l_i) - \q_i \Y (\l_i)\, ] \Bigr ) ,
\ee{expval}
where we write $\D (\l_i, \q_i)= \prod_{i<j} (\l_i -\l_j -
\q_i\q_j)\, $ for the measure  .
We introduce the one--superloop correlators

\be
\widehat{W}(p \mid\, ) = N \,\Blangle \,\sum_i \frac{\q_i}{p-\l_i} \,
\Brangle
\qquad
\mbox{and}
\quad
\widehat{W}(\, \mid p) = N \, \Blangle \,\sum_i \frac{1}{p-\l_i}\, \Brangle ,
\ee{Wwidehat}
which act as generating functionals  for the one--point correlators
$\langle \,\sum_i {\l_i}^k \,\rangle$ and $\langle \,\sum_i \q_i {\l_i}^k\,
\rangle$
upon expansion in $p$.  This easily generalizes to higher--point correlators
with the $(n|m)$--superloop correlator

\bea
\lefteqn{\widehat{W}(p_1,\ldots ,p_n \mid q_1,\ldots ,q_m) =}
  \label{Wwidehatnm}\\  & & N^{n+m} \,
\Blangle \,
\sum_{i_1} \frac{\q_{i_1}}{p_1-\l_{i_1}} \, \ldots \, \sum_{i_n}
\frac{\q_{i_n}}{p_n-\l_{i_n}} \, \sum_{j_1}\frac{1}{q_1-\l_{j_1}} \,
\ldots\, \sum_{j_m} \frac{1}{q_m-\l_{j_m}} \,\Brangle . \nonumber
\eea
Quite analogously to the bosonic case \cite{Amb} these correlators may be
obtained from
the partition function ${\cal Z}$ by application of the superloop insertion
operators $\d/ \d V(p)$ and  $\d/ \d \Y (p)$:

\be
\widehat{W}(p_1,\ldots ,p_n \mid q_1,\ldots ,q_m)= \frac{1}{\cal Z}
\dY{p_1} \,\ldots\,
\dY{p_n}\, \dV{q_1}\,\ldots\, \dV{q_m}
\,\, {\cal Z},
\ee{Wwidehatnm2}
where

\be
\dV{p}= -\sum_{k=0}^\infty\frac{1}{p^{k+1}} \frac{\pa}{\pa g_k}
 \quad\mbox{and}\quad
\dY{p}= -\sum_{k=0}^\infty\frac{1}{p^{k+1}} \frac{\pa}{\pa \x_{k+1/2}}.
\ee{superloopensertionops}
In particular equation \refer{Wwidehat} can now be written as
\mbox{$\widehat{W}(\, \mid p)= \d\, \ln {\cal Z} /\d V
(p)$} and \mbox{$\widehat{W}(p\mid\,)=
\d\, \ln {\cal Z}/ \d \Y (p)$}.

However, it is convenient to work with the connected part of the
$(n|m)$--superloop correlators, denoted by $W$. They may
be obtained from the free energy \mbox{$F=N^{-2}\,\ln\,{\cal Z}$} through

\bea
\lefteqn{W(p_1,\ldots ,p_n \mid q_1,\ldots ,q_m)=
\dY{p_1} \,\ldots\,
\dY{p_n}\, \dV{q_1}\,\ldots\, \dV{q_m}
\,\,  F}\label{Wnm}\\ & = &
N^{n+m-2} \,
\Blangle \,
\sum_{i_1} \frac{\q_{i_1}}{p_1-\l_{i_1}} \, \ldots \, \sum_{i_n}
\frac{\q_{i_n}}{p_n-\l_{i_n}} \, \sum_{j_1}\frac{1}{q_1-\l_{j_1}} \,
\ldots\, \sum_{j_m} \frac{1}{q_m-\l_{j_m}} \,\Brangle _C .
\nonumber\eea
Note that  \mat{n\leq 2} due to the structure of $F$ mentioned below.

With the normalizations chosen above, one assumes that these correlators
enjoy the genus expansion \cite{Alv2}

\be
W(p_1,\ldots ,p_n \mid q_1,\ldots ,q_m)= \sum_{g=0}^\infty \, \frac{1}{N^{2g}}
\, W_g (p_1,\ldots ,p_n \mid q_1,\ldots ,q_m).
\ee{genusexpWnm}
Similarly one has the genus expansion

\be
F= \sum_{g=0}^\infty \, \frac{1}{N^{2g}} \, F_g
\ee{genusexpF}
for the free energy.

\subsection{Superloop Equations}

The superloop equations of our model are two Schwinger--Dyson equations, which
we derive in appendix A. They were
first stated in \cite{Alv,Alv2}, and we present them in an integral form for
the loop correlators $W(p\mid \, )$ and $W(\, \mid p )$.

The Grassmann--odd superloop equation reads

\be
\cint{\w}{C} \, \frac{V^\prime (\w)}{p-\w}\, W (\w \mid\, ) \, + \,
\cint{\w}{C} \, \frac{\Y (\w)}{p-\w} \, W (\, \mid \w) =
W(p \mid \, ) \, W(\,\mid p) \, + \, \frac{1}{N^2} \, W(p \mid p)
\ee{superloop1}
and its counterpart, the Grassmann--even superloop equation, takes the form

$$
\cint{\w}{C} \, \frac{ V^\prime (\w ) \, W(\, \mid \w ) \, + \,
\Y ^\prime (\w ) \,
W(\w \mid \, )}{p - \w} \, -\, \frac{1}{2} \, \frac{d}{dp}\, \cint{\w}{C} \,
\frac{\Y (\w )\, W( \w \mid )\, }{p-\w} = \phantom{W(p \mid \, )\,}$$
 \be
 \frac{1}{2} \, \Bigl [ \, W(\, \mid p)^2 \, - \, W(p \mid \, )\, W^\prime (p
 \mid \, ) \, + \, \frac{1}{N^2}\, \Bigl ( \, W(\, \mid p,p) \, - \,
\frac{d}{dq}\,
 W( p,q \mid \, ) \,  \Bigr | _{p=q} \, \Bigr ) \Bigr ] .
\ee{superloop2}
In the derivation we have assumed that the loop correlators have one--cut
structure,
i.e.\ in the limit $N\ra \infty$ we assume that the eigenvalues are contained
in a finite interval \mbox{$[x,y]$}.
Moreover $C$ is a curve around the cut.

Note the similarity to the loop equations for the hermitian matrix model
\cite{Mak}.
The equations \refer{superloop1} and \refer{superloop2}
are equivalent to the superloop equations discussed in
\cite{Alv,Alv2}, which may be seen by performing the contour
integrals and taking the residues at $\w =p$ and $\w = \infty$. Moreover eqs.\
\refer{superloop1} and \refer{superloop2} simply encode the
super--Virasoro constraints $G_{k-1/2}\,
{\cal Z}=0$ and $L_k\, {\cal Z} = 0$ for $k \geq 0$, which the model obeys by
construction.

The key to the solution of these complicated equations order by order
in $N^{-2}$
 is the observation made in \cite{Bec,McA} that
the free energy $F$ depends at most quadratically on fermionic coupling
constants. Via eq. \refer{Wnm} this directly translates to the one-loop
correlators, which we from now on write as

\bea
W(p \mid \, ) &=& v(p) \\
W(\, \mid p)  &=& u(p) \, + \, \widehat{u}(p).
\eea
Here $v(p)$ is of order one in fermionic couplings, whereas $u(p)$ is taken
to be of order zero and $\widehat{u} (p)$ of order two in the fermionic
coupling
constants $\x_{k+1/2}$. This
observation allows us to split up the two superloop equations
\refer{superloop1}
and \refer{superloop2} into a set of four equations, sorted by their order in
the $\x_{k+1/2}$'s. Doing this we obtain

\medskip
\leftline{Order 0:}
\bea
\lefteqn{
\cint{\w}{C} \, \frac{V^\prime (\w )}{p-\w} \, u(\w ) = \frac{1}{2}\, u(p)^2
\, + \, \frac{1}{2}\, \frac{1}{N^2}\, \dV{p}\, u(p) \, - \, \frac{1}{2}\,
\frac{1}{N^2}
\, \frac{d}{dq}\, \dY{p} \, v(q) \Bigr | _{p=q}} \nonumber \\ & &
\phantom{\cint{\w}{C} \, \frac{V^\prime (\w )}{p-\w} \, u(\w ) =
\frac{1}{2}\, u(p)^2
\, + \, \frac{1}{2}\, \frac{1}{N^2}\, \dV{p}\, u(p) \, - \,
\frac{1}{2}\, \frac{1}{N^2}
\, \frac{d}{dq}\, \dY{p} \, v(q) \Bigr | _{p=q}}
\la{order0}\eea
\leftline{Order 1:}
\bea\lefteqn{
\cint{\w}{C}\, \frac{V^\prime (\w )}{p-\w }\, v(\w )\, + \, \cint{\w}{C} \,
\frac{\Y (\w )}{p-\w}\, u(\w ) = v(p)\, u(p) \, + \, \frac{1}{N^2}\,
\dV{p}\, v(p)}
\nonumber \\ & &
\phantom{\cint{\w}{C} \, \frac{V^\prime (\w )}{p-\w} \, u(\w ) =
\frac{1}{2}\, u(p)^2
\, + \, \frac{1}{2}\, \frac{1}{N^2}\, \dV{p}\, u(p) \, - \, \frac{1}{2}\,
\frac{1}{N^2}
\, \frac{d}{dq}\, \dY{p} \, v(q) \Bigr | _{p=q}}
\la{order1}\eea
\leftline{Order 2:}
\bea\lefteqn{
\cint{\w}{C} \, \frac{V^\prime (\w )}{p-\w }\, \widehat{u}(\w ) \, + \,
\cint{\w}{C} \, \frac{\Y^\prime (\w )}{p-\w}\, v(\w ) \, - \, \frac{1}{2}\,
\frac{d}{dp} \, \cint{\w}{C}\,  \frac{\Y (\w )}{p-\w }\, v(\w ) =
\phantom{\frac{d}{dp}\, \cint{\w}{C}}} \nonumber \\ & &
 u(p)\, \widehat{u}(p)\, -\, \frac{1}{2}\, v(p)\, \frac{d}{dp}\, v(p) \, + \,
\frac{1}{2}\, \frac{1}{N^2}\, \dV{p}\, \widehat{u}(p) \nonumber \\ & &
\phantom{\cint{\w}{C} \, \frac{V^\prime (\w )}{p-\w} \, u(\w ) =
\frac{1}{2}\, u(p)^2
\, + \, \frac{1}{2}\, \frac{1}{N^2}\, \dV{p}\, u(p) \, - \, \frac{1}{2}\,
\frac{1}{N^2}
\, \frac{d}{dq}\, \dY{p} \, v(q) \Bigr | _{p=q}}
\la{order2}\eea
\leftline{Order 3:}
\bea\lefteqn{
\cint{\w}{C}\, \frac{\Y (\w )}{p-\w }\, \widehat{u} (\w ) = v(p)\, \widehat{u}
(p).} \nonumber \\ & &
\phantom{\cint{\w}{C} \, \frac{V^\prime (\w )}{p-\w} \, u(\w ) =
\frac{1}{2}\, u(p)^2
\, + \, \frac{1}{2}\, \frac{1}{N^2}\, \dV{p}\, u(p) \, - \, \frac{1}{2}\,
 \frac{1}{N^2}
\, \frac{d}{dq}\, \dY{p} \, v(q) \Bigr | _{p=q}}
\la{order3}\eea
It is the remarkable form of these four equations which allows us to develop
an iterative procedure to determine $u_g(p), v_g(p), \widehat{u}_g(p)$
and $F_g$
genus by genus. Plugging the genus expansions into these equations lets them
decouple partially, in the sense that the equation of order 0 at genus $g$
only involves
$u_g$ and lower genera contributions. The order 1 equation then only contains
$v_g$, $u_g$ and lower genera results and so on. The first thing to do,
however,
is to find the solution for $g=0$.

\sect{The Planar Solution}

In the following the planar solution for the superloop correlators is given for
a general potential. It was first obtained in \cite{Alv2}. We present
it in a very compact integral form augmented by the use of new
variables to characterize the potentials, the moments.

\subsection{Solution for $u_0(p)$ and $v_0(p)$}

In the limit $N\ra \infty$ the order 0 equation \refer{order0} becomes

\be
\cint{\w}{C}\, \frac{V^\prime (\w )}{p-\w }\, u_0(\w )\,
= \, \frac{1}{2}\, u_0(\w )^2.
\ee{order0genus0}
This equation is well known, as up to a factor of $1/2$ it is
nothing but the planar
loop equation of the
hermitian matrix model. With the above assumptions on the one--cut structure
and by demanding that $u(p)$ behaves as $1/p$ for $p\ra \infty$ one
finds \cite{Mig}

\be
u_0 (p)\, = \, \cint{\w}{C} \, \frac{V^\prime (\w )}{p-\w}\, \biggl [ \,
\frac{(p-x)(p-y)}{(\w -x)(\w -y)}\, \biggr ]^{1/2} ,
\ee{u_0}
where the endpoints  $x$ and $y$ of the cut on the real axis
are determined by the
following requirements:

\be
0=\cint{\w}{C}\, \frac {V^\prime(\w )}{\sqrt{(\w -x)(\w -y)}} , \quad \quad
1=\cint{\w}{C}\, \frac{\w\, V^\prime(\w )}{\sqrt{(\w -x)(\w -y)}},
\ee{determinexy}
deduced from our knowledge that $W(\, \mid p)= 1/p + {\cal O}(p^{-2})$.

The order 1 equation \refer{order1} in the $N\ra \infty $ limit determining
the odd loop correlator $v_0(p)$ reads

\be
\cint{\w}{C}\, \frac{V^\prime (\w )}{p-\w} \, v_0 (\w ) \, + \, \cint{\w}{C}\,
\frac{\Y (\w )}{p-\w}\, u_0(\w ) \, =\, v_0(p)\, u_0 (p).
\ee{order1genus0}
It is solved by

\be
v_0(p) \, =\, \cint{\w}{C}\,
\frac{\Y (\w )}{p-\w}\, \biggl [ \, \frac{(\w -x)(\w -y)}
{(p-x)(p-y)} \, \biggr ]^{1/2} \,\, + \,\, \frac{\c}{\sqrt{(p-x)(p-y)}} .
\ee{v_0}
Here $\c$ is a constant not determined by eq.\ \refer{order1genus0}, in fact
$\c=N^{-1}\, \langle \, \sum_i \q_i \, \rangle$ in the planar limit.
It will be
determined in the analysis of the two remaining equations \refer{order2}
and \refer{order3}. We verify the above solution in appendix B.

\subsection{Moments and Basis Functions}

Let us now define new variables characterizing the
potentials $V(p)$ and $\Y (p)$. Instead of the couplings $g_k$
we introduce the bosonic moments $M_k$ and $J_k$ defined by \cite{Amb}

\bea
M_k &=& \cint{\w}{C}\, \frac{V^\prime (\w )}{ (\w - x)^k} \,
\frac{1}{[\, (\w -x)\, (\w-y)\, ]^{1/2}}
, \quad k\geq 1 \\ &&\nonumber\\
J_k &=& \cint{\w}{C}\, \frac{V^\prime (\w )}{ (\w - y)^k} \,
\frac{1}{[\, (\w -x)\, (\w-y)\, ]^{1/2}}
, \quad k\geq 1 ,
\eea
and the couplings $\x_{k+1/2}$ are replaced by the fermionic moments

\bea
\X_k &=& \cint{\w}{C}
\, \frac{\Y (\w )}{(\w -x)^k}\, [\,(\w -x)(\w -y)\, ]^{1/2}
,\quad k\geq 1 \\ &&\nonumber\\
\L_k &=& \cint{\w}{C}\,
\frac{\Y (\w )}{(\w -y)^k}\, [\,(\w -x)(\w -y)\, ]^{1/2}
,\quad k\geq 1 .
\eea
The main motivation for introducing these new variables is that, for each term
in the genus expansion of the free energy and the correlators, the dependence
on an
infinite number of coupling constants arranges itself nicely into a function
of a {\it finite} number of moments.

We further introduce the basis functions $\c ^{(n)}(p)$
and $\Y^{(n)}(p)$ recursively

\bea
\c ^{(n)}(p) &= &
\frac{1}{M_1}\, \Bigl ( \, \f^{(n)}_x(p)\, - \,\sum_{k=1}^{n-1} \c^{(k)}
(p)\, M_{n-k+1}\, \Bigr ) , \\
& &\nonumber \\
\Y^{(n)}(p) &= &\frac{1}{J_1}\,
\Bigl ( \, \f^{(n)}_y(p)\, - \,\sum_{k=1}^{n-1} \Y^{(k)}
(p)\, J_{n-k+1} \Bigr ) ,
\eea
where

\bea
\f^{(n)}_x  (p) & = & (p-x)^{-n}\, [\, (p-x)(p-y)\, ]^{-1/2}, \\
& & \nonumber \\
\f^{(n)}_y  (p) & = & (p-y)^{-n}\, [\, (p-x)(p-y)\, ]^{-1/2},
\eea
following \cite{Amb}.

It is easy to show that for the linear operator $\widehat{V}^\prime$ defined by
\be
\widehat{V}^\prime \, f(p) = \cint{\w}{C}\, \frac{V^\prime (\w )}{p-\w}\,
f(\w )
\ee{opVprime}
and appearing in the superloop equations we have

\bea
\Bigl ( \,
\widehat{V^\prime} - u_0(p) \, \Bigr ) \, \c ^{(n)}(p) &=& \frac{1}{(p-x)
^n}, \quad n\geq 1,  \label{VprimeChi} \\
& & \nonumber \\
\Bigl ( \,
\widehat{V^\prime} - u_0(p) \, \Bigr ) \, \Y^{(n)}(p) &=& \frac{1}{(p-y)
^n}, \quad n\geq 1. \label{VprimePsi}
\eea
Moreover, $\f^{(0)}_x=\f^{(0)}_y\equiv \f^{(0)}$ lies in the kernel of $(\,
\widehat{V^\prime} - u_0(p)\, )$.

\subsection{Solution for $\widehat{u}_0$ and $\c$}

Next consider the order 2 equation \refer{order2} at genus 0

\bea
\Bigl ( \, \widehat{V^\prime} - u_0(p) \, \Bigr ) \, \widehat{u}_0 &=&
\frac{1}{2}\,
\frac{d}{dp}\,\cint{\w}{C}\, \frac{\Y (\w )}{p-\w}\, v_0(\w ) \, -\,
\cint{\w}{C}\, \frac{\Y^\prime (\w )}{p-\w }\, v_0(\w ) \nonumber \\
& & - \frac{1}{2}\, v_0(p)\, \frac{d}{dp}\, v_0(p) .\label{order2genus0}
\eea
Plugging  eq.\
\refer{v_0} into the right hand side of this equation yields after a
somewhat lengthy calculation

\be
\Bigl ( \, \widehat{V^\prime} - u_0(p) \, \Bigr ) \, \widehat{u}_0 \, =\,
\frac{1}{2}\, \frac{\X_2\, (\X_1 -\c )}{(x-y)}\, \frac{1}{p-x}\, - \,
\frac{1}{2}\, \frac{\L_2\, (\L_1 -\c )}{(x-y)}\, \frac{1}{p-y}.
\ee{page20g}
With eqs.
\refer{VprimeChi} and \refer{VprimePsi} this immediately tells us that

\be
\widehat{u}_0(p) = \frac{1}{2}\, \frac{\X_2\, (\X_1 -\c )}{(x-y)}
\,  \c ^{(1)}(p) \,- \,
\frac{1}{2}\, \frac{\L_2\, (\L_1 -\c )}{(x-y)}\,\Y^{(1)}(p).
\ee{uhat0first}
There can be no contributions proportional to the zero mode $\f^{(0)}(p)$,
as we know that $\widehat{u}(p)$ behaves as
${\cal O}(p^{-2})$ for $p \ra \infty$.

Finally we determine the odd constant $\c$. This is done by employing the
\hbox{order 3} equation for $g=0$, i.e.

\be
\cint{\w}{C}\, \frac{\Y (\w )}{p-\w}\,\widehat{u}_0(\w ) \, -\, v_0 (p)\,
\widehat{u}_0(p)=0.
\ee{order3genus0}
After insertion of eqs.\ \refer{v_0} and \refer{uhat0first} one can show that

\be
0=\Bigl (\, \widehat{\Y} \, -\, v_0(p)\, \Bigr )\, \widehat{u}_0(p) =
\frac{1}{2}\,
\frac{ \X_2\, (\X_1 -\c )\, (\L_1 -\c )}{M_1\, (x-y)^{3}\, (p-y)}\,
-\, \frac{1}{2}\,
\frac{ \L_2\, (\L_1 -\c )\, (\X_1 -\c )}{J_1\, (x-y)^{3}\, (p-x)}
\ee{cresult}
where we have defined the operator $\widehat{\Y}$ by

\be
\widehat{\Y}\, f(p) = \cint{\w}{C}\, \frac{\Y (\w )}{p-\w}\, f(\w) ,
\ee{Psiop}
in accordance to $\widehat{V^\prime}$.
The result \refer{cresult} lets us finally read off the coefficient $\c$ as

\be
\c \, = \, \frac{1}{2}\, (\, \X_1 \, + \, \L_1\, ).
\ee{c}

Putting it all
together, we may now write down the complete genus 0 solution for
the one--superloop correlators $W(\, \mid p)$ and $W(p\mid \,)$:

\bea
W_0(\,\mid p) &=&\cint{\w}{C}\, \frac{V^\prime (\w )}{p-\w}\,
\biggl [ \, \frac{(p-x)(p-y)}{(\w -x)(\w -y)}\, \biggr ]^{1/2}\,
\phantom{+\,
\frac{1}{4}\, \frac{\X_2\, \XL}{M_1\, (x-y)}\, \f^{(1)}_x(p)} \label{W0p_0}\\
& &
+\,
\frac{1}{4}\, \frac{\X_2\, \XL}{M_1\, (x-y)}\, \f^{(1)}_x(p)
+\, \frac{1}{4}\, \frac{\L_2\, \XL}{J_1\, (x-y)}\, \f^{(1)}_y(p)\nonumber\\
&&\nonumber\\&&\nonumber\\
W_0(p\mid \, ) &=& \cint{\w}{C}\, \frac{\Y (\w )}{p-\w}\, \biggl [\,
\frac{(\w -x)(\w -y)}{(p-x)(p-y)}\, \biggr ] ^{1/2} \, +\, \frac{1}{2}\,
\frac{\X_1\, +\, \L_1}{[\, (p-x)(p-y)\, ]^{1/2}}.\nonumber \\ & &
\label{Wp0_0}
\eea
One can show that this solution is equivalent to the one obtained by
Alvarez--Gaum\' e, Becker, Becker, Emparan and Ma\~ nes in \cite{Alv2}.

\sect{The Iterative Procedure}

Our iterative solution of the superloop equations results in a certain
representation of the free energy and the loop correlators in terms of the
moments and basis functions defined in section 3.2. We will show that it
suffices to know $u_g(p)$ and $v_g(p)$ only up to a zero mode in order to
calculate $F_g$. We give explicit results for genus one.

\subsection{The Iteration for $u_g$ and $v_g$}

The correlators $u_g(p)$ and $v_g(p)$ are determined by the order 0 and order 1
equations \refer{order0} and \refer{order1} after insertion of the genus
expansions \refer{genusexpWnm} of these operators. We find

\bea
\Vop \, u_g(p) &=&
\sum_{g^\prime=1}^{g-1}\, u_{g^\prime}(p)\, u_{g-g^\prime}(p)
\, +\, \frac{1}{2}\, \dV{p}\, u_{g-1}(p)\, \nonumber \\ & &
-\, \frac{1}{2}\, \frac{d}{dq}\,\dY{p}\,
v_{g-1}(q)\Bigr | _{p=q}
\label{order0genusg}
\eea
and

\bea
\Vop \, v_g(p)  &=& -
\Yop\, u_g(p)\, +\,  \sum_{g^\prime=1}^{g-1}\, v_{g^\prime}(p)
\, u_{g-g^\prime}(p) \, \nonumber \\ & &
+\, \dV{p}\, v_{g-1}(p)
\label{order1genusg}
\eea
at genus $g\geq 1$.
{}From the structure of these equations we directly deduce that
$u_g(p)$ and $v_g(p)$ will be linear combinations of the basis functions
$\c^{(n)}(p)$
and $\Y^{(n)}(p)$. By eqs.\ \refer{VprimeChi} and \refer{VprimePsi}
the coefficients of this linear
combination may be read off the poles $(p-x)^{-k}$ and $(p-y)^{-k}$ of
the right hand sides of eqs. \refer{order0genusg}
and \refer{order1genusg} after a partial fraction decomposition.

Let us demonstrate how this works for $g=1$.
According to eq.\ \refer{order0genusg} for $u_1(p)$
we first calculate $\d\, u_0/\d V(p) $. We then need to know the derivatives
$\d\, x/\d V(p)$ and $\d\, y/ \d V(p)$. They can be obtained from eq.\
\refer{determinexy} and read

\be
\frac{\d \, x}{\d V(p)}= \frac{1}{M_1}\, \f^{(1)}_x (p), \quad
\frac{\d \, y}{\d V(p)}= \frac{1}{J_1}\, \f^{(1)}_y (p).
\ee{dVxy}
Using the relation

\be
\dV{p}\, V^\prime (\w ) = \frac{d}{dp}\, \frac{1}{p-\w}
\ee{dVVprime}
one finds

\be
\dV{p}\, u_0(p) = \frac{1}{8}\, \px{2} + \frac{1}{8}\,\py{2}
- \frac{1}{4\, d}\, \px{} + \frac{1}{4\, d}\, \py{},
\ee{dVu0}
where $d= x-y$.

Next we determine $\d\, v_0(q)/\d \Y (p)$.  Using the relation

\be
\dY{p}\, \Y (q) = -\, \frac{1}{p-q}
\ee{dYY}
and the result

\bea
\frac{\d\,
\X_k}{\d \Y (p)} &=& \d_{k1}\, -\,  \frac{[\, (p-x)(p-y)\, ]^{1/2}}{ (p-x)^k}
\label{dYX} \\ &&\nonumber \\
\frac{\d\, \L_k}{\d\Y (p)}
&=& \d_{k1}\, - \, \frac{[\, (p-x)(p-y)\, ]^{1/2}}{ (p-y)^k}
\label{dYL}
\eea
for $k \geq 1$, one finds

\be
\frac{d}{dq}\, \dY{p}\, v_0(q)\Bigr |_{p=q} = - \dV{p} \, u_0 (p).
\ee{dqdYv0}
This enables us to write down $u_1(p)$,

\be
u_1 (p) = \frac{1}{8}\, \c^{(2)} (p) + \frac{1}{8}\, \Y^{(2)}(p)
-\frac{1}{4\, d}\, \c^{(1)} +\frac{1}{4\, d}\, \Y^{(1)}(p).
\ee{u1}
Note that up to the overall factor of two this is identical to the one--loop
correlator of
the hermitian matrix model \cite{Amb}, as it has to be \cite{Bec}.

Now we solve eq.\
\refer{order1genusg} at $g=1$ for $v_1(p)$. It is important to
realize that generally eq.\ \refer{order1genusg} fixes $v_g(p)$ only up to a
zero mode
contribution $\k_g\, \f^{(0)}(p)$. This comes from the fact that, unlike
for the bosonic $u(p)$, we do not know the coefficient of the $p^{-1}$ term for
$v(p)$. The zero mode coefficient $\k_g$ will be fixed by  requiring
$v_g(p)$ to be a total derivative of the free energy $F_g$.

In order to calculate $\d v_0/ \d V(p)$ we make use of the relation

\be
\frac{\d\, \X_k}{\d V(p)} = (k-\frac{1}{2})\, \X_{k+1}\,
\frac{1}{M_1}\, \f^{(1)}_x (p)
 +\frac{1}{2}\, \Bigl [ \, \sum_{r=2}^k \frac{\X_r}{(-d)^{1+k-r}} +
\frac{\X_1 -\L_1}{(-d)^k} \, \Bigr ]\, \frac{1}{J_1}\, \f^{(1)}_y (p),
\ee{dVXi}
as well as $\d\, \L_k/\d V(p)$ obtained from the above by the replacements
$x \lra y$, $M_k \lra J_k$, $\X_k \lra \L_k$ and $d \ra -d$.
The derivatives $\d\, M_k/\d V(p)$ and $\d\, J_k/\d V(p)$ were calculated
in \cite{Amb}

\bea
\frac{\d \, M_k}{\d V(p)} &=& -\frac{1}{2}\, (p-x)^{-k-1/2}(p-y)^{-3/2} -
                (k+1/2)\, \f^{(k+1)}_x (p) \nonumber \\
&& + \frac{1}{2}\,
\Bigl [ \, \frac{1}{(-d)^k} - \sum_{i=1}^k \frac{1}{(-d)^{k-i+1}} \,
\frac{M_i}{J_1}\, \Bigr ] \, \f^{(1)}_y (p)
\nonumber \\
& & + (k+1/2)\, \frac{M_{k+1}}{M_1}\, \f^{(1)}_x (p), \label{dMdV}
\eea
and $\d\, J_k/\d V(p)$ is obtained by the usual replacements. Using these
and the earlier results one has

\bea
\frac{\d\, v_0}{\d V(p)}
 = W_0(p\mid p) &=& \bra{ - \frac{\XL}{4\, d\, M_1}}\, \px{3} \, +\,
  \bra{- \frac{\XL}{4\, d\, J_1} }\, \py{3} \nonumber\\ & + &
  \bra{\frac{\X_2}{4\, d\, M_1}}\, \px{2}\,
  +\,\bra{- \frac{\L_2}{4\, d\, J_1}}\, \py{2}  \nonumber\\ & + &
  \bra{\frac{\L_2}{4\, d^2\, J_1}- \frac{\X_2}{4\, d^2\, M_1}}\, \px{}
  \nonumber \\ &+&
  \bra{\frac{\X_2}{4\, d^2\, M_1} -
  \frac{\L_2}{4\, d^2\, J_1}}\, \py{}. \nonumber\\
  &&\label{Wpp}
\eea

For the evaluation of the right hand side of eq.\ \refer{order1genusg} at
genus $g$
we also need to know how the operator $(\, \widehat{\Y}- v_0(p)\, )$ acts on
the functions $\f^{(n)}_x(p)$ and $\f^{(n)}_y(p)$,
in terms of which $u_g(p)$ is given. A straightforward calculation yields

\bea
\Yop \, \f^{(n)}_x (p) &=&
\sum_{k=1}^{n+1}\, \px{k}\, \biggl [ \, -\frac{\XL}{2\,
(-d)^{n+2-k}} \, -\, \sum_{l=2}^{n+2-k} \frac{\X_l}{(-d)^{n+3-k-l}} \, \biggr ]
\nonumber \\
& &+\, \py{}\, \biggl[ \, -\frac{\XL}{2\, (-d)^{n+1}}\, \biggr ] \label{Yopfxn}
\eea
as well as the
analogous expression for $(\, \widehat{\Y}- v_0(p)\, )\,\f^{(n)}_y(p)$
obtained from eq.\ \refer{Yopfxn} by the replacements
$x \lra y$, $M_k \lra J_k$, $\X_k \lra \L_k$ and $d \ra -d$.

We now have collected
all the ingredients needed to evaluate the right hand side
of eq.\ \refer{order1genusg}. After a partial fraction decomposition we
may read off the poles at $x$ and $y$, and therefore obtain the coefficients
of the linear combination in the basis functions. We arrived at
the result for $g=1$ with the aid of {\it Maple}, namely

\be
v_1(p) =\sum_{i=1}^3\Bigl ( \, B^{(i)}_1\, \c^{(i)}(p) \, +\, E^{(i)}_1\,
\Y^{(i)}(p)\, \Bigr ) \, + \, \k_1\, \f^{(0)}(p),
\ee{v1}
where the coefficients $B^{(i)}_1$ and $E^{(i)}_1$ are given by

\bea
B^{(1)}_1 & = & -\frac{1}{8}\, \frac{\X_3}{d\, M_1} +
\frac{1}{8}\, \frac{\X_2}{d^2\, M_1} + \frac{1}{4}\,\frac{\L_2}{d^2\, J_1}
\nonumber \\
& & +\frac{1}{8}\,\frac{M_2\, \X_2}{d\, {M_1}^2}
-\frac{1}{16}\,\frac{M_2\,\XL}{d^2\, {M_1}^2} +\frac{1}{16}\,\frac{J_2\,\XL}
{d^2\, {J_1}^2} \nonumber \\
& & -\frac{3}{16}\,
\frac{\XL}{d^3\, M_1} - \frac{3}{16}\, \frac{\XL}{d^3\, J_1},
\nonumber \\ &&\nonumber\\
B^{(2)}_1 &=&
\frac{1}{8}\,\frac{\X_2}{d\, M_1} + \frac{1}{16}\,\frac{M_2\, \XL}
{d\, {M_1}^2} + \frac{3}{16}\,\frac{\XL}{d^2\, M_1}, \nonumber \\
&& \nonumber \\
B^{(3)}_1 &=& -\frac{5}{16}\,\frac{\XL}{d\, M_1}, \label{v1coeffs}
\eea
and \mat{E^{(i)}_1 = B^{(i)}_1 ( M\lra J, \X \lra \L, d\ra -d)}.

Yet $\k_1$ is still undetermined. To compute it and the remaining doubly
fermionic part $\widehat{u}_g(p)$ of the loop correlator $W_g(\,\mid p)$ one
can employ the order 2 and order 3 eqs.\ \refer{order2} and \refer{order3}
at genus $g$. It is, however, much easier to construct the free energy $F_g$ at
this stage from our knowledge of $u_g(p)$ and $v_g(p)$.

\subsection{The Computation of $F_g$ and $\k_g$}

As mentioned earlier, the free energy of the supereigenvalue model depends
at most quadratically on the fermionic coupling constants.
In this subsection we present an algorithm which
allows us to determine $F_g$ and $\k_g$ as soon as the results for $u_g(p)$ and
$v_g(p)$ (up to the zero mode coefficient $\k_g$) are known.

One can show \cite{Bec} that the purely bosonic part of the free energy
$F_g$ is just twice the free energy of the hermitian matrix model. By using the
results of Ambj\o rn {\it et al.\ }\cite{Amb} one may then compute the
bosonic part of $F_g$ from $u_g(p)$.

The strategy for the part of $F_g$ quadratic in fermionic couplings
consists in rewriting
$v_g(p)$ as a total derivative in the fermionic potential
$\Y (p)$. We know that eqs.\ \refer{dYX} and \refer{dYL} imply

\be
\frac{\d\,\XL}{\d\Y (p)} = - d\, \f^{(0)}(p)
\ee{dYXL}
and

\be
\frac{\d\, \X_k}{\d\Y (p)} = -\,\vq^{(k)}_x(p), \quad
\frac{\d\,\L_k}{\d\Y (p)}=-\vq^{(k)}_y(p),\quad
k\geq 2
\ee{dYXkLk}
where

\bea
\vq^{(k)}_x(p)&=&
(p-x)^{-k}\, [\, (p-x)(p-y)\, ]^{1/2} \quad k \geq 1, \nonumber\\
& &\label{vfdef}\\
\vq^{(k)}_y(p)&=&
(p-y)^{-k}\, [\, (p-x)(p-y)\, ]^{1/2} \quad k \geq 1. \nonumber
\eea

Let us again specialize to \mat{g=1}.
Using the above we can reexpress eq.\ \refer{v1} as

\bea\lefteqn{
\dY{p}\, F_1 \, +\, \k_1 \, \frac{1}{d}\, \dY{p}\,\XL = \sum_{i=1}^3
\Bigl ( \, B^{(i)}_1\, \c^{(i)}(p) \, +\, E^{(i)}_1\, \Y^{(i)}(p)\, \Bigr )
\phantom{\XL\XL} }\nonumber\\
&=& \sum_{r=2}^4 \Bigl ( \, \b^{(r)}_1\, \vq^{(r)}_x(p) \, +\, \e^{(r)}_1\,
\vq^{(r)}_x(p)\, \Bigr )\, +\, \g_1\, [\, \vq^{(1)}_x(p)-\vq^{(1)}_y(p)\, ],
\label{p49}
\eea
with the new coefficients $\b^{(r)}_1$, $\e^{(r)}_1$ and $\g_1$ completely
determined by the known coefficients $B^{(i)}_1$
and $E^{(i)}_1$. As the new functions
$\vq^{(r)}_x(p)$ and $\vq^{(r)}_y(p)$ are total derivatives in $\Y (p)$, this
equation allows us to calculate $\k_1$ and $F_1$.

With the help of {\it Maple} the zero mode coefficient $\k_1$
of $v_1(p)$ becomes

\bea
\k_1 &=&
{\frac {11\,\X_{{2}}}{16\,{d}^{3}{M_{{1}}}^{2}}}
-{\frac {11\,\L_{{2}}}{16\,{d}^{3}{J_{{1}}}^{2}}}
+{\frac {5\,\L_{{2}}J_{{2}}}{8\,{d}^{2}{J_{{1}}}^{3}}}
+{\frac {5\,\X_{{2}}M_{{2}}}{8\,{d}^{2}{M_{{1}}}^{3}}}
-{\frac {5\,\X_{{2}}M_{{3}}}{16\,d{M_{{1}}}^{3}}} \nonumber \\ & &
+{\frac {5\,\L_{{2}}J_{{3}}}{16\,d{J_{{1}}}^{3}}}
-{\frac {\X_{{2}}}{16\,{d}^{3}J_{{1}}M_{{1}}}}
+{\frac {\L_{{2}}}{16\,{d}^{3}J_{{1}}M_{{1}}}}
-{\frac {\X_{{2}}J_{{2}}}{16\,{d}^{2}{J_{{1}}}^{2}M_{{1}}}}
-{\frac {\L_{{2}}M_{{2}}}{16\,{d}^{2}{M_{{1}}}^{2}J_{{1}}}} \nonumber \\ & &
+{\frac {3\,\X_{{2}}{M_{{2}}}^{2}}{8\, d{M_{{1}}}^{4}}}
-{\frac {3\,\L_{{2}}{J_{{2}}}^{2}}{8\,d{J_{{1}}}^{4}}}
+{\frac {3\,\L_{{3}}J_{{2}}}{8\,d{J_{{1}}}^{3}}}
-{\frac {3\,\X_{{3}}M_{{2}}}{8\,d{M_{{1}}}^{3}}}
+{\frac {5\,\X_{{4}}}{16\,d{M_{{1}}}^{2}}}\nonumber \\ & &
-{\frac {5\,\L_{{4}}}{16\,d{J_{{1}}}^{2}}}
-{\frac {5\,\X_{{3}}}{8\,{d}^{2}{M_{{1}}}^{2}}}
-{\frac {5\,\L_{{3}}}{8\,{d}^{2}{J_{{1}}}^{2}}}
+{\XL}\,
\Bigl \{ {\frac {3\,{J_{{2}}}^{2}}{8\,{d}^{2}{J_{{1}}}^{4}}} \nonumber \\ & &
     -{\frac {3\,{M_{{2}}}^{2}}{8\,{d}^{2}{M_{{1}}}^{4}}}
     -{\frac {5\,M_{{2}}}{8\,{d}^{3}{M_{{1}}}^{3}}}
     -{\frac {5\,J_{{2}}}{8\,{d}^{3}{J_{{1}}}^{3}}}
     +{\frac {5\,M_{{3}}}{16\,{d}^{2}{M_{{1}}}^{3}}}
     -{\frac {5\,J_{{3}}}{16\,{d}^{2}{J_{{1}}}^{3}}} \nonumber \\ & &
     +{\frac {11}{16\,{d}^{4}{J_{{1}}}^{2}}}
     -{\frac {11}{16\,{d}^{4}{M_{{1}}}^{2}}}
     +{\frac {M_{{2}}}{16\,{d}^{3}J_{{1}}{M_{{1}}}^{2}}}
     +{\frac {J_{{2}}}{16\,{d}^{3}{J_{{1}}}^{2}M_{{1}}}} \Bigr \}
\label{kappa}
\eea
and the doubly fermionic part of $F_1$ is constructed as well.

The result for the free energy at genus 1 then reads

\bea
F_1 &=& -\frac{1}{12}\,
\ln\, M_1 \,-\,\frac{1}{12}\,\ln\, J_1\, -\,\frac{1}{3}\,
\ln\, d \nonumber \\ & &
-\XL\, \Bigl \{ \,
  {\frac {11\,\X_{{2}}}{16\,{d}^{4}\,{M_{{1}}}^{2}}}
-{\frac {11\,\L_{{2}}}{16\,{d}^{4}\,{J_{{1}}}^{2}}}
+{\frac {5\,\L_{{2}}\, J_{{2}}}{8\,{d}^{3}\,{J_{{1}}}^{3}}}
+{\frac {5\,\X_{{2}}\, M_{{2}}}{8\,{d}^{3}\,{M_{{1}}}^{3}}} \nonumber \\ & &
-{\frac {5\,\X_{{2}}\, M_{{3}}}{16\,d^2\,{M_{{1}}}^{3}}}
+{\frac {5\,\L_{{2}}\, J_{{3}}}{16\,d^2\,{J_{{1}}}^{3}}}
-{\frac {\X_{{2}}\, J_{{2}}}{16\,{d}^{3}\,{J_{{1}}}^{2}M_{{1}}}}
-{\frac {\L_{{2}}\, M_{{2}}}{16\,{d}^{3}\, {M_{{1}}}^{2}J_{{1}}}}
\nonumber \\ & &
+{\frac {3\,\X_{{2}}\,{M_{{2}}}^{2}}{8\,d^2\,{M_{{1}}}^{4}}}
-{\frac {3\,\L_{{2}}\,{J_{{2}}}^{2}}{8\,d^2\,{J_{{1}}}^{4}}}
+{\frac {3\,\L_{{3}}\, J_{{2}}}{8\,d^2\,{J_{{1}}}^{3}}}
-{\frac {3\,\X_{{3}}\, M_{{2}}}{8\,d^2\,{M_{{1}}}^{3}}}
+{\frac {5\,\X_{{4}}}{16\,d^2\,{M_{{1}}}^{2}}}\nonumber \\ & &
-{\frac {5\,\L_{{4}}}{16\,d^2\,{J_{{1}}}^{2}}}
-{\frac {\X_{{3}}}{2\,{d}^{3}\,{M_{{1}}}^{2}}}
-{\frac {\L_{{3}}}{2\,{d}^{3}\,{J_{{1}}}^{2}}}
+{\frac {3\,\X_{{2}}}{16\,{d}^{4}\, J_{{1}}M_{{1}}}}
-{\frac {3\,\L_{{2}}}{16\,{d}^{4}\, J_{{1}}M_{{1}}}}\Bigr \}\nonumber \\ & &
+{\frac {\X_{2}\, \X_{{3}}}{8\,{d}^{2}\, {M_{{1}}}^{2}}}
+{\frac {\L_2\, \L_{{3}}}{8\,{d}^{2}\,{J_{{1}}}^{2}}}
-{\frac {\X_2\,\L_2}{4\,{d}^{3}\, J_{{1}}M_{{1}}}}.
\label{F1}
\eea

\subsection{The Iteration for $\widehat{u}_g(p)$}

The remaining part of the loop correlator $W(\,\mid p)$ is now easily derived
from $F_g$ by applying the loop insertion operator $\d/\d V(p)$
to its doubly fermionic part.

For genus 1 the result is

\be
\widehat{u}_1(p)= \sum_{i=1}^4
\Bigl ( \, \widehat{A}^{(i)}_1\, \c^{(i)}(p) \, +\,
\widehat{D}^{(i)}_1\, \Y^{(i)}(p)\, \Bigr ) ,
\ee{uhat1}
where

\bea
\widehat{A}^{(4)}_1 &=& -\frac{35\, \XL\, \X_2}{32\, d^2\, {M_1}^2}
\nonumber \\ & &\nonumber \\
\widehat{A}^{(3)}_1 &=& \XL\, \Bigl \{ \,
                -\frac{15\,  \X_3}{16\, d^2\,{M_1}^2} +
                \frac{45\,\X_2}{32\, d^3\, {M_1}^2} +
                \frac{25\,\X_2\, M_2}{32\, d^2\, {M_1}^3} -
                \frac{5\,\L_2}{32\, d^3\, J_1\, M_1} \, \Bigr \}
\nonumber \\ & & \nonumber \\
\widehat{A}^{(2)}_1 &=&
{\frac {21\,\XL\,\X_{{3}}}{16\,{d}^{3}{M_{{1}}}^{2}}}
-{\frac {3\,\XL\,\X_{{2}}{M_{{2}}}^{2}}{8\,{d}^{2}{M_{{1}}}^{4}}}
+{\frac {3\,\XL\,\X_{{2}}J_{{2}}}{32\,{d}^{3}{J_{{1}}}^{2}M_{{1}}}}
+{\frac {\X_{{2}}\,\X_{{3}}}{8\,{d}^{2}{M_{{1}}}^{2}}} \nonumber \\ & &
+{\frac {3\,\L_{{2}}\,\X_{{2}}}{8\,J_{{1}}M_{{1}}{d}^{3}}}
+{\frac {3\,\XL\,\X_{{3}}M_{{2}}}{4\,{d}^{2}{M_{{1}}}^{3}}}
-{\frac {15\,\XL\,\X_{{4}}}{16\,{d}^{2}{M_{{1}}}^{2}}}
-{\frac {\X_3\,\X_{{2}}}{4\,{d}^{2}{M_{{1}}}^{2}}} \nonumber \\ & &
-{\frac {51\,\XL\,\X_{{2}}}{32\,{d}^{4}{M_{{1}}}^{2}}}
-{\frac {\L_{{2}}\,\XL\, M_{{2}}}{32\,J_{{1}}{d}^{3}{M_{{1}}}^{2}}}
-{\frac {\L_2\,\XL}{4\,J_{{1}}{d}^{4}M_{{1}}}}\nonumber \\ & &
-{\frac {33\,\XL\,\X_{{2}}M_{{2}}}{32\,{d}^{3}{M_{{1}}}^{3}}}
+{\frac {5\,\XL\,\X_{{2}}M_{{3}}}{16\,{d}^{2}{M_{{1}}}^{3}}}
-{\frac {9\,\XL\,\X_{{2}}}{32\,J_{{1}}{d}^{4}M_{{1}}}}
\nonumber \\ & & \nonumber \\
\widehat{A}^{(1)}_1 &=&
{\frac {33\,\XL\,\X_{{2}}M_{{2}}}{32\,{d}^{4}{M_{{1}}}^{3}}}
-{\frac {\XL\,\X_{{2}}M_{{2}}}{32\,{d}^{4}J_{{1}}{M_{{1}}}^{2}}}
-{\frac {35\,\XL\,\X_{{5}}}{32\,{d}^{2}{M_{{1}}}^{2}}} \nonumber \\ & &
-{\frac {3\,\XL\,\X_{{2}}{J_{{2}}}^{2}}{16\,{d}^{3}{J_{{1}}}^{4}}}
+{\frac {15\,\XL\,\X_{{4}}M_{{2}}}{16\,{d}^{2}{M_{{1}}}^{3}}}
+{\frac {45\,\XL\,\X_{{4}}}{32\,{d}^{3}{M_{{1}}}^{2}}}\nonumber \\ & &
+{\frac {5\,\XL\,\X_{{2}}}{16\,{d}^{5}J_{{1}}M_{{1}}}}
-{\frac {7\,\XL\,\L_{{3}}}{32\,{d}^{4}{J_{{1}}}^{2}}}
-{\frac {\X_2\,\L_{{3}}}{16\,{d}^{3}{J_{{1}}}^{2}}}
+{\frac {\X_3\,\X_{{2}}}{16\,{d}^{3}{M_{{1}}}^{2}}} \nonumber \\ & &
+{\frac {51\,\XL\,\X_{{2}}}{32\,{d}^{5}{M_{{1}}}^{2}}}
-{\frac {51\,\XL\,\X_{{3}}}{32\,{d}^{4}{M_{{1}}}^{2}}}
-{\frac {\X_2\,\X_{{3}}}{8\,{d}^{3}{M_{{1}}}^{2}}}
+{\frac {5\,\X_2\,\X_{{4}}}{16\,{d}^{2}{M_{{1}}}^{2}}}\nonumber \\ & &
-{\frac {\L_2\,\X_{{2}}}{2\,J_{{1}}M_{{1}}{d}^{4}}}
-{\frac {7\,\L_2\,\XL\, J_{{2}}}{32\,{J_{{1}}}^{3}{d}^{4}}}
+{\frac {3\,\L_2\,\X_{{3}}}{8\,{d}^{3}J_{{1}}M_{{1}}}}
+{\frac {7\,\L_2\,\XL}{16\,{J_{{1}}}^{2}{d}^{5}}}\nonumber \\ & &
+{\frac {5\,\XL\,\X_{{2}}\, J_{{2}}}{16\,{J_{{1}}}^{3}{d}^{4}}}
-{\frac {11\,\XL\,\X_{{2}}}{32\,{J_{{1}}}^{2}{d}^{5}}}
-{\frac {5\,\XL\,\X_{{2}}M_{{3}}}{16\,{d}^{3}{M_{{1}}}^{3}}}  \nonumber \\ & &
-{\frac {9\,\XL\,\X_{{3}}{M_{{2}}}^{2}}{16\,{d}^{2}{M_{{1}}}^{4}}}
+{\frac {3\,\XL\,\X_{{3}}J_{{2}}}{32\,{d}^{3}{J_{{1}}}^{2}M_{{1}}}}
-{\frac {9\,\XL\,\X_{{3}}}{32\,J_{{1}}M_{{1}}{d}^{4}}}\nonumber \\ & &
+{\frac {15\,\XL\,\X_{{3}}M_{{3}}}{32\,{d}^{2}{M_{{1}}}^{3}}}
-{\frac {9\,\XL\,\X_{{3}}M_{{2}}}{8\,{d}^{3}{M_{{1}}}^{3}}}
+{\frac {5\,\XL\,\X_{{2}}J_{{3}}}{32\,{d}^{3}{J_{{1}}}^{3}}}\nonumber \\ & &
-\frac{\X_2}{2\, d}\,\biggl ( \,
 -{\frac {11\,\L_{{2}}}{16\,{d}^{3}{J_{{1}}}^{2}}}
+{\frac {5\,\L_{{2}}J_{{2}}}{8\,{d}^{2}{J_{{1}}}^{3}}}
+{\frac {5\,\L_{{2}}J_{{3}}}{16\,d{J_{{1}}}^{3}}}
+{\frac {\L_{{2}}}{16\,{d}^{3}J_{{1}}M_{{1}}}}\nonumber \\ & &
-{\frac {\L_{{2}}M_{{2}}}{16\,{d}^{2}{M_{{1}}}^{2}J_{{1}}}}
-{\frac {3\,\L_{{2}}{J_{{2}}}^{2}}{8\,d{J_{{1}}}^{4}}}
+{\frac {3\,\L_{{3}}J_{{2}}}{8\,d{J_{{1}}}^{3}}}
-{\frac {3\,\X_{{3}}M_{{2}}}{8\,d{M_{{1}}}^{3}}}
+{\frac {5\,\X_{{4}}}{16\,d{M_{{1}}}^{2}}}\nonumber \\ & &
-{\frac {5\,\L_{{4}}}{16\,d{J_{{1}}}^{2}}}
-{\frac {5\,\X_{{3}}}{8\,{d}^{2}{M_{{1}}}^{2}}}
-{\frac {5\,\L_{{3}}}{8\,{d}^{2}{J_{{1}}}^{2}}}
+{\frac {3\,\XL\,\X_{{2}}{M_{{2}}}^{2}}{8\,{d}^{3}{M_{{1}}}^{4}}}
\nonumber \\ & &
-{\frac {\XL\,\X_{{2}}J_{{2}}}{8\,{d}^{4}{J_{{1}}}^{2}M_{{1}}}}
+{\frac {\L_2\,\XL\, M_{{2}}}{16\,{d}^{4}J_{{1}}{M_1}^{2}}}
+{\frac {11\,\L_2\,\XL}{32\,J_{{1}}{d}^{5}M_{{1}}}} \nonumber \\ & &
+\XL\, \Bigl \{\,
     {\frac {3\,{J_{{2}}}^{2}}{8\,{d}^{2}{J_{{1}}}^{4}}}
-{\frac {3\,{M_{{2}}}^{2}}{8\,{d}^{2}{M_{{1}}}^{4}}}
-{\frac {5\,M_{{2}}}{8\,{d}^{3}{M_{{1}}}^{3}}}
-{\frac {5\,J_{{2}}}{8\,{d}^{3}{J_{{1}}}^{3}}} \nonumber \\ & &
+{\frac {5\,M_{{3}}}{16\,{d}^{2}{M_{{1}}}^{3}}}
-{\frac {5\,J_{{3}}}{16\,{d}^{2}{J_{{1}}}^{3}}}
+{\frac {11}{16\,{d}^{4}{J_{{1}}}^{2}}}
-{\frac {11}{16\,{d}^{4}{M_{{1}}}^{2}}} \nonumber \\ & &
+{\frac {M_{{2}}}{16\,{d}^{3}J_{{1}}{M_{{1}}}^{2}}}
+{\frac {J_{{2}}}{16\,{d}^{3}{J_{{1}}}^{2}M_{{1}}}}
  \, \Bigr \} \,\biggr ),
\label{uhat1coeff}
\eea
and the analogue expressions for the $\widehat{D}^{(i)}_1$ obtained from the
above by replacing $M\lra J$, $\X\lra\L$ and $d\ra -d$.

\subsection{General Structure of $u_g$, $v_g$, $\widehat{u}_g$ and $F_g$}

In the following subsection we find
the number of moments and basis functions the
free energy and the superloop correlators at genus $g$ depend on. For this
write \Fbg for the purely bosonic and \Ffg for the doubly fermionic
parts of the free energy.

Ambj\o rn {\it et al.\ }\cite{Amb, Amb2} have shown that the free energy of the
hermitian matrix model depends on $2(3g-2)$ moments. This directly translates
to $F_g^{\mbox{\scriptsize bos}}$.
Similarly as \mat{u_g=\d\, F^{\mbox{\scriptsize bos}}_g/\d  V(p)} and
with eq.\ \refer{dMdV} we see that $u_g$
contains \mat{2(3g-1)} bosonic moments
and basis functions up to order \mat{(3g-1)}, i.e.

\be
u_g(p)= \sum_{k=1}^{3g-1}\, A^{(k)}_g\, \c^{(k)}(p) \, +\,
D^{(k)}_g\, \Y^{(k)}(p).
\ee{structureug}
For the structure of $v_g$ consider the leading--order poles on the RHS of
eq.\ \refer{order1genusg}. Label this order by $n_g$, then
with eqs.\ \refer{dVXi}, \refer{dMdV}
and \refer{Yopfxn} the three terms on the RHS
of eq.\ \refer{order1genusg} give rise to the following poles of leading
order

\bea
\Yop \, u_g(p) & : & (3g-1) + 1 \nonumber \\
v_{g^\prime}\, u_{g-g^\prime} & : & n_{g^\prime} + \Bigl (\, 3 \,
(g-g^\prime) -1\,
\Bigr ) \, +\, 1 \nonumber \\
\dV{p}\, v_{g-1} & : & n_{g-1} \, +\, 3 .\label{ng}
\eea
{}From the above we deduce that \mat{n_g=3\, g}, and therefore

\be
v_g(p)= \sum_{k=1}^{3g}\, B^{(k)}_g\, \c^{(k)}(p) \, +\, E^{(k)}_g\,
\Y^{(k)}(p).
\ee{structurevg}
As the highest bosonic moments in $v_g$ come from the highest--order
basis functions, we see that $v_g$ depends on $2(3\, g)$ bosonic moments.
To find the dependence on
the number of fermionic moments recall eqs. \refer{dYX} and \refer{dYL}. In
order for $v_g$ to have a leading contribution of \mat{\c^{(3g)}(p)} the
fermionic part of the free energy \Ffg must contain
$\X_{3g+1}$. We are thus led to the conclusion that \Ffg and $v_g$
both depend on \mat{2(3\, g+1)} fermionic moments. As the application of the
loop insertion operator \mat{\d /\d\Y (p)} does not change the number of
bosonic moments, \Ffg must contain \mat{2(3\, g)} bosonic moments.

Knowing the structure of \Ffg then tells us with eqs. \refer{dVXi} and
\refer{dMdV}
that $\widehat{u}_g$ depends on \mat{2(3\, g +1)} bosonic and \mat{2(3\, g+2)}
fermionic moments. For genus $g$ it reads

\be
\widehat{u}_g (p) = \sum_{k=1}^{3g+1} \, \widehat{A}^{(k)}_g \, \c^{(k)}(p)
\, + \, \widehat{D}^{(k)}_g\, \Y^{(k)}(p).
\ee{strustureuhatg}
The genus $g$ contribution to the \mat{(\, |s)}--superloop correlator
\mat{W_g(\, |p,\ldots ,p)} will then depend on \mat{2(3\, g +s)} bosonic
and \mat{2(3\, g+s+1)} fermionic moments. Similarly the genus $g$
\mat{(1|s)}--superloop correlator \mat{W_g(p|p,\ldots ,p)} is a function
of \mat{2(3\, g + s)} bosonic and \mat{(2(3\, g+s+1)} fermionic moments.

This concludes our analysis of the iterative process.

\sect{Conclusions}

We have studied the supereigenvalue model away from the double scaling limit.
The superloop correlators of this model obey a set of integral equations, the
superloop equations. These two equations could be split up into a set of four
equations, sorted by their order in fermionic coupling constants. By a change
of variables from coupling constants to moments we were
able to present the planar solution of the superloop equations
for general potentials in a very compact form.
The remarkable structure of the superloop equations enabled us to develop an
iterative procedure for the calculation of higher--genera contributions to the
free energy and the superloop correlators. Here it proved sufficient to solve
the two lowest--order equations at genus $g$ for
the purely bosonic $u_g(p)$ and
the fermionic $v_g(p)$ (up to a zero mode contribution). The zero mode
as well as the doubly fermionic part of the free energy could then be
found by rewriting $v_g(p)$ as a total derivative in the fermionic potential.
The purely bosonic part of the free energy can be calculated
with the methods of \cite{Amb}.
In principle the application of loop insertion operators to the free
energy then yields
arbitrary multi--superloop correlators. As we demonstrated for genus one,
in practice these expressions become quite lengthy. We ended with a survey
on the general structure of the free energy and the multi--superloop
correlators for genus $g$.

We believe that this paper describes a suitable approach to the solution of the
supereigenvalue model. The structures of the iterative
solutions to the superloop
equations and to the loop equations of the hermitian matrix model show
interesting relations. This may give new
hope for finding a generalized matrix--based formulation
of the supereigenvalue model and thus allowing a geometrical
interpretation of it. Moreover we think that our
results underline the effectiveness of the iterative scheme of Ambj\o rn
{\it et al.}, which should be applicable to further models of similar
type as well.

We expect that the described iterative solution for the supereigenvalue model
may also be set up in the double scaling limit.
This, however, is left for future work.

\bigskip\noindent
\underline{Acknowledgements:} I wish to thank P.\ Adamietz, G.\
Akemann, L.\ Alvarez-Gaum\'e, J.\ Ambj\o rn and O.\ Lechtenfeld
for useful discussions.

\begin{appendix}
\pagebreak
\section*{Appendix}
\renewcommand{\theequation}{A.\arabic{equation}}
\setcounter{equation}{0}
\subsection*{A.  Derivation of the Superloop Equations}

In order to derive the first of the two superloop equations for the model
of eq.\ \refer{model} consider the shift in integration variables

\be
\l_i \ra \l_i + \q_i \frac{\e}{p-\l_i} \quad \mbox{and} \quad \,\,\,
\q_i \ra \q_i + \frac{\e}{p-\l_i}
\ee{shift1}
where $\e$ is an odd constant. Under these we find that

\be
\prod_i d\l_i\, d\q_i \ra (1-\e\sum_i \frac{\q_i}{(p-\l_i)^2})\,
\prod_i d\l_i\,
\q_i
\ee{}
and the measure transforms as

\be
\D (\l_i,\q_i) \ra (1-\e \sum_{i\neq j} \frac{\q_i}{(p-\l_i)(p-\l_j)})\,
\D (\l_i,\q_i).
\ee{}
The vanishing of the terms proportional to $\e$ then gives us
the Schwinger--Dyson equation

\be
\langle \, N \Bigl\{ \sum_i \frac{1}{p-\l_i} \, \Bigr ( V^\prime (\l_i) \q_i +
\Y (\l_i) \Bigr ) \Bigr \}  \, - \,\sum_i \frac{\q_i}{p-\l_i} \, \sum_j
\frac{1}{p-\l_j} \, \rangle = 0.
\ee{Schwinger-Dyson-1}
Note that $\langle \,\sum_i \q_i\,(p-\l_i)^{-1} \,
\sum_j (p-\l_j)^{-1} \, \rangle = N^{-2}\, \widehat{W} (p\mid p)$
with the above definitions. In order to
transform eq.\ \refer{Schwinger-Dyson-1} into an
integral equation we define the
bosonic and fermionic density operators

\be
\r (\l ) = \frac{1}{N} \, \sum_i \langle \, \d (\l-\l_i ) \, \rangle
\quad \mbox{and} \quad \,\,\,
r (\l)= \frac{1}{N} \, \sum_i \langle \, \q_i \, \d (\l-\l_i ) \, \rangle .
\ee{density-ops}
With these the first sum in eq.\ \refer{Schwinger-Dyson-1} may be written as

$$
N^2\, \int d\l \, \Bigl ( \, r(\l)\, \frac{V^\prime (\l)}{p-\l}\,+\,
\r (\l)\, \frac{\Y (\l)}{p-\l}\, \Bigr ) =\phantom{N^2\, \int d \l
\,\, r(\l )\, \Bigl
[ \cint{\w}{C} \, \frac{1}{\w -\l}\,  \frac{V^\prime (\w)}{p-\w}
\, \Bigr ] + }$$
 \be
 N^2\, \int d \l \,\, r(\l )\, \Bigl [ \cint{\w}{C}
 \, \frac{1}{\w -\l}\,  \frac{V^\prime (\w)}{p-\w} \, \Bigr ] +
   N^2\, \int d \l \, \,\r (\l )\, \Bigl [ \cint{\w}{C} \, \frac{1}{\w -\l}\,
 \frac{\Y (\w)}{p-\w} \, \Bigr ]
 \ee{calc2.2}
 where we assume that the real eigenvalues $\l_i$
 are contained within a finite
 interval \mbox{$x \, <
\l_i \, < y \, ,\, \forall i$}.  Moreover $C$ is a curve
 around $0$ with radius
\mbox{$R > \max (|x|,|y|)$} and we choose \mbox{$|p|> R$}.
 Performing the $\l$
integrals in eq. \refer{calc2.2} gives us the one--superloop
 correlators. The full Schwinger--Dyson equation \refer{Schwinger-Dyson-1}
 may then be expressed in the integral form

\be
\cint{\w}{C} \, \frac{V^\prime (\w)}{p-\w}\, \widehat{W} (\w \mid\, ) \, + \,
\cint{\w}{C} \, \frac{\Y (\w)}{p-\w} \, \widehat{W} (\, \mid \w) = N^{-2}\,
\widehat{W} (p \mid p) .
\ee{superloop1hat}
Rewriting this in terms of the connected superloop correlators $W$ yields eq.
\refer{superloop1}.

The derivation of the second superloop equation goes along the same lines by
performing the shift

\be
\l_i \ra \l_i + \frac{\e}{p-\l_i} \quad \mbox{and} \quad \,\,\,
\q_i \ra \q_i + \frac{1}{2}\, \frac{\e \, \q_i}{(p-\l_i)^2 }
\ee{shift2}
with $\e$ even and infinitesimal. Similar steps as the ones discussed above
then lead us to the second superloop equation

$$
\cint{\w}{C} \, \frac{V^\prime (\w )\, \widehat{W} (\, \mid \w ) \, +\,
\Y^\prime (\w) \,
\widehat{W}(\w \mid \, )}{p-\w} \, -\, \frac{1}{2} \,
\frac{d}{dp}\, \cint{\w}{C}\,
\frac{\Y (\w )\, \widehat{W}(\w \mid \, )}{p-\w}=
$$
\be
\frac{1}{2}\, \frac{1}{N^2}\, \widehat{W}(\, \mid p,p) \, - \, \frac{1}{2}\,
\frac{1}{N^2}\, \frac{d}{dq}\, \widehat{W}(p,q \mid \, ) \Bigr | _{p=q}
\ee{superloophat2}
which, after rephrasing in connected quantities, gives eq.\ \refer{superloop2}.

\renewcommand{\theequation}{B.\arabic{equation}}
\setcounter{equation}{0}
\subsection*{B. Proof of the Solution $v_0 (p)$ for Eq.\ \refer{order1genus0}}

Using eqs. \refer{u_0} and \refer{v_0} the right hand side of
eq. \refer{order1genus0} reads

\bea
u_0(p)\, v_0(p) &=& \cint{\w}{C_1}\,\cint{z}{C_2}\, \frac{V^\prime(\w)\,\Y (z)}
{(p-\w )(p-z)}\, \biggl [ \frac{(z-x)(z-y)}{(\w -x)(\w -y)}\, \biggr ]^{1/2}
\nonumber\\
& &
+ \cint{\w}{C_1}\,
\frac{V^\prime(\w )}{p-\w}\, \frac{\c}{[(\w -x)(\w -y)]^{1/2}},
\label{B1}
\eea
and the left hand becomes

$$
\cint{\w}{C_1}\, \cint{z}{C_2}\, \frac{V^\prime(\w )\, \Y (z)}{(p-\w)(\w -z)}\,
\biggl [ \frac{(z-x)(z-y)}{(\w -x)(\w -y)}\biggr ]^{1/2}  \,
\phantom{\cint{\w}{C_1}\,\cint{\w}{C_1}\,\cint{\w}{C_1}\,} $$
$$+ \,
\cint{\w}{C_1}\,\frac{V^\prime (\w )}{p-\w}\, \frac{\c}{[(\w -x)(\w -y)]^{1/2}}
\phantom{\biggl [ \frac{(\w -x)(\w -y)}{(z-x)(z-y)} \biggr]^{1/2}.}
$$
\be\phantom{\cint{\w}{C_1}\, \cint{z}{C_2}}
+\, \cint{\w}{C_1}\,\cint{z}{C_2}\, \frac{\Y (\w )\, V^\prime (z)}{(p-\w )
(\w -z)}\, \biggl [ \frac{(\w -x)(\w -y)}{(z-x)(z-y)} \biggr]^{1/2}.
\ee{B2}
Now in the last term pull the contour integral $C_2$  over the curve $C_1$. One
can show that the contribution from the extra pole vanishes. After renaming
$\w \leftrightarrow z$ and combining the first and third terms one gets
eq.\ \refer{B1}. Thus eq.\ \refer{order1genus0} is verified.
\end{appendix}
\pagebreak

\end{document}